\documentclass[runningheads]{llncs}
\usepackage{graphicx}
\usepackage{amsfonts}
\usepackage{amsmath}
\usepackage{multirow}
\usepackage{wrapfig}
\usepackage{appendix}
\begin{document}
\title{Multi-scale GANs for Memory-efficient Generation of High Resolution Medical Images}
\titlerunning{Multi-scale GANs for High Resolution Medical Images}
%
\author{Hristina Uzunova, Jan Ehrhardt, Fabian Jacob, Alex Frydrychowicz, Heinz Handels}

\authorrunning{H. Uzunova, et. al.}

%
\institute{Institute of Medical Informatics, University of L\"ubeck}
\maketitle              
\begin{abstract}
Currently generative adversarial networks (GANs) are rarely applied to medical images of large sizes, especially 3D volumes, due to their large computational demand. We propose a novel multi-scale patch-based GAN approach to generate large high resolution 2D and 3D images. Our key idea is to first learn a low-resolution version of the image and then generate patches of successively growing resolutions conditioned on previous scales. In a domain translation use-case scenario,  3D thorax CTs of size $512^3$ and thorax X-rays of size $2048^2$ are generated and we show that, due to the constant GPU memory demand of our method, arbitrarily large images of high resolution can be generated. Moreover, compared to common patch-based approaches, our multi-resolution scheme enables better image quality and prevents patch artifacts.
\end{abstract}
\section{Introduction}
Generative adversarial networks (GANs) \cite{GANs} have shown impressive results for photo-realistic image synthesis in the last couple of years \cite{pix2pix,progressive,stackGAN}. They also offer numerous applications in medical image analysis, such as generating images for data augmentation, image reconstruction and image synthesis or domain adaptation.
Despite the undeniable success and the large variety of applications, GANs still struggle to generate images of high resolution. A reason for that is the fact that generated images are easier to distinguish from real ones at higher resolutions, which hinders the training process \cite{progressive}. Further reasons are computational demands and memory requirements of current network architectures. To deal with the first issue, Karras et al. \cite{progressive} recently proposed a progressive learning strategy for GANs that starts with low resolution and adds finer details throughout the training. 
Using this strategy, the authors were able to generate high resolution 2D images of size $1024\times1024$. However, the training process already requires $\sim$16 GB of GPU RAM, which means that larger images can only be generated with special and expensive hardware. The problem is further aggravated in 3D imaging typically demanded in clinical routine. An early attempt for the 3D application of GANs is made in \cite{3dgan}, where furniture shapes of size $64^3$ are generated using a DCGAN architecture. Even though the task of generating only shapes (no texture or intensities) is fairly simple, its computational demands are borderline to most consumer-class GPUs.

The image size limitations make GANs impracticable for many medical image applications, e.g. for thoracic CTs, brain MRIs, or high resolution X-ray images. 
In \cite{3DMedGAN1} the authors claim to be forced to use only half of the image size ($128\times128\times 54$) due to GPU memory limits of the dedicated hardware (NVIDIA DGX system). The largest 3D output size of a GAN found in literature is $128^3$ \cite{3DMedGAN2}, however no memory requirements are mentioned. An usual approach to overcome such computational challenges is slice-/patch-wise generation \cite{SPIE3DGAN,deepmedic}. Unfortunately, those methods suffer from artifacts between patches/slices due to noncontinuous transitions. Dealing with this problem by applying patch overlaps and averaging, leads to blurry results and loss of image detail. The intuition behind patch inconsistencies is that when patches are generated independently, they do not have any global intensity information. Thus, \cite{deepmedic} proposes to additionally observe a larger area around each patch of the input image to cope with this issue. 
Even though shown to be well suitable for segmentation and would probably improve patch artifacts in strictly paired image translation (e.g. CT to MR), such an approach cannot be applied to image generation from scratch (or sparsely conditioned), and its effect is limited when the image size drastically exceeds the chosen patch size.

In this work we propose a memory-efficient multi-scale GAN approach for the generation of high-resolution medical images in high quality. Our approach combines a progressive multi-scale learning strategy with a patch-wise approach, where low-resolution image content is learned first, and image patches at higher resolutions are conditioned on the previous scales to preserve global intensity information. We demonstrate the ability to generate realistic large images on thoracic X-rays of size $2048^2$ and 3D lung CTs of size $512^3$. 
Further, we show that w.r.t. the growing side length of an isotropic 3D image, the memory requirements for popular GANs grow cubical, while they stay constant for any image size using our approach.
Although the presented method is theoretically suitable for from-scratch image generation, in this work, we apply a conditional GAN for unsupervised domain adaptation. This application features topology preserving style transfer, where in contrast to strictly paired image translation, our approach does not assume corresponding images from two domains and enables the possibility to translate arbitrary datasets to a desired domain. We show that image translation from different CT reconstruction kernels, different devices and different acquisition parameters to a particular desired domain is possible, keeping the original 3D image size.
The generated images are evaluated with respect to a known ground-truth image emphasizing the quality of generated images.
\section{Methods}
  GANs are generative models that learn to map a random noise vector $\mathbf{z}$ to an output image $y$ using a generator function $G : \mathbf{z}\rightarrow y$ \cite{GANs}. An extension of regular GANs are the conditional GANs, that learn the mapping from an observed image $x$ additionally, $G : \{x,\mathbf{z}\}\rightarrow y$. To ensure that the generator produces realistically looking images that cannot be distinguished from real ones, an adversarial discriminator $D$ is enclosed in the training process, aiming to perfectly distinguish between real images and generator's fakes. 
 \begin{figure}[t]
   \centering
     \includegraphics[width=0.99\textwidth]{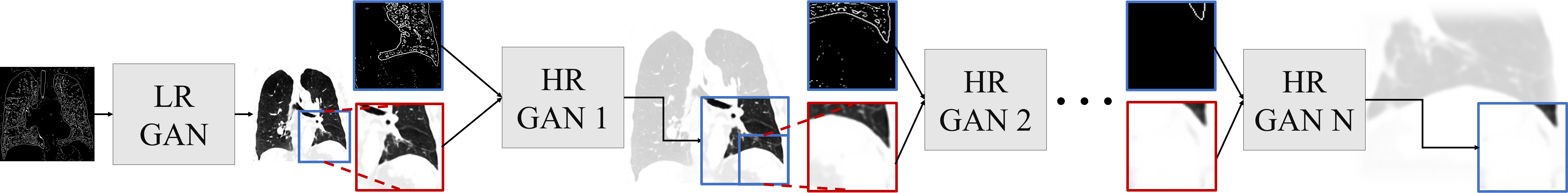}
     \caption{An overview of our method. Generate the whole image with a low resolution (LR) GAN, then subsequently increase the resolution by generating patches with multiple high resolution (HR) GANs conditioned on the previous scales.  Blue: patches of original resolution for the current scale; red: upscaled patches of lower resolution.}
     \label{fig:overview}
\end{figure} 
\subsection{Multi-scale Conditional GANs}
The idea of multi-scale image generation to obtain high resolutions, is inspired by early works like \cite{laplace}, where, based on a Laplacian pyramid, multiple GANs for increasing resolutions are used, and more recent approaches \cite{stackGAN} that achieve photo-realistic image quality by iterative resolution refinement applying successive GANs. However, those methods still assume that at a certain stage the whole full-resolution image is propagated through the network, requiring extremely high GPU capacities especially for 3D images.  
Here, we propose to overcome this issue by first generating the whole image in a low resolution, then iteratively proceed by generating image patches of constant size but 
growing resolutions. Since the GANs for each scale are trained separately, a particular maximum size is never exceeded during training, so the GPU demand stays constantly low, facilitating the generation of arbitrarily large images. Further, we enable a consistent and realistic look of the generated images by conditioning each high resolution scale on the previous lower resolution image, thus providing patches with global intensity information and preventing patch artifacts and inconsistencies (see Fig.~\ref{fig:overview} for a method overview).
 A further benefit of progressively learning increasing resolutions, is that each GAN has a rather simple task to learn compared to generating the whole high-resolution image from scratch. This does not only enable a simpler training process, but also ensures sharp results for each scale and thus a consistent high-resolution result. 
 
We consider an additional condition, namely the object edges, since our use-case scenario implies topology preserving image domain translation. In the lowest resolution (LR) GAN, the whole low-resolution edge image is used as an input. In the higher-resolution (HR) GANs, two conditions are used: a patch from the image of the previous resolution, upscaled to the size of the current scale; and a patch from the edge image of the current scale.
Note, that the patch size of each resolution is constant, however, as the overall image becomes larger after each HR GAN, the relative image fraction per patch decreases, while resolution grows. 
The objective of this learning process can be described as follows. For multiple conditional images $x_0\dots x_n$ with resolutions $0\dots n$, output images $y_0 \dots y_n$, where $y_n$ represents the final output image, are generated using separate generators $G_{0\dots n}$ and discriminators $D_{0\dots n}$ with the objectives
\begin{equation*}
\begin{aligned}
    \mathcal{L}_{cGAN}(G_0,D_0)&=\\
   & \mathbb{E}_{x_0,y_0}[\log D_0(x_0,y_0)]+\mathbb{E}_{x_0,\mathbf{z}}[\log (1-D_0(x_0,G_0(x_0,\mathbf{z})))]\\
    \mathcal{L}_{cGAN}(G_i,D_i)&=\mathbb{E}_{x_{p_i},y_{p_i}}[\log D_i(x_{p_i},y_{p_{i-1}},y_{p_i})]+\\
  & \mathbb{E}_{x_{p_i},\mathbf{z}}[\log (1-D_i(x_{p_i},y_{p_{i-1}},G_i(x_{p_i},y_{p_{i-1}},\mathbf{z})))],
    \end{aligned}
\end{equation*}
where $x_{p_i}$ and $y_{p_i}$ are patches of the conditional image $x_i$ and the generated image $y_i$ respectively, with $i \in [1,n]$.

\subsection{Architecture and Training}
Different generator's architectures were chosen for $G_0$ and $G_{1\dots n}$, since the tasks of generating whole low resolution images and high resolution patches differ in a variety of requirements. The LR GAN  uses a U-Net architecture, which is able to filter out many unimportant details and generalize better due to its bottleneck. Its tendency to result in more blurry images is negligible in the context of low-resolution images.
For the patch generation by the HR GANs, ResNet blocks are chosen, since they are known to produce sharp results by keeping the input image resolution unchanged. The higher overfitting risk of not having a bottleneck is diminished due to the stronger conditioning (on the previous scales) and the overall large number of patches compared to the number of images used.  For the discriminators $D_{0\dots n}$ a regular fully-convolutional architecture is chosen\footnote{The complete architecture is available in the supplemental materials.}. 

Data augmentation is crucial to our method, because cascading approaches are prone to propagate errors up from lower resolutions. To deal with this issue we corrupt a percentage of the low resolution images while training each HR GAN: we apply random noise, Gaussian blurring and vary the image resolution, also to make the patch generation less dependent on perfect edge extraction, the edge images are perturbed with noise. In our experience those techniques help to overcome overadapting to either of the inputs.

\section{Results and Experiments}
\subsection{Memory Requirements for 3D GANs}
\begin{figure}[t]
\centering
     \includegraphics[width=0.6\textwidth]{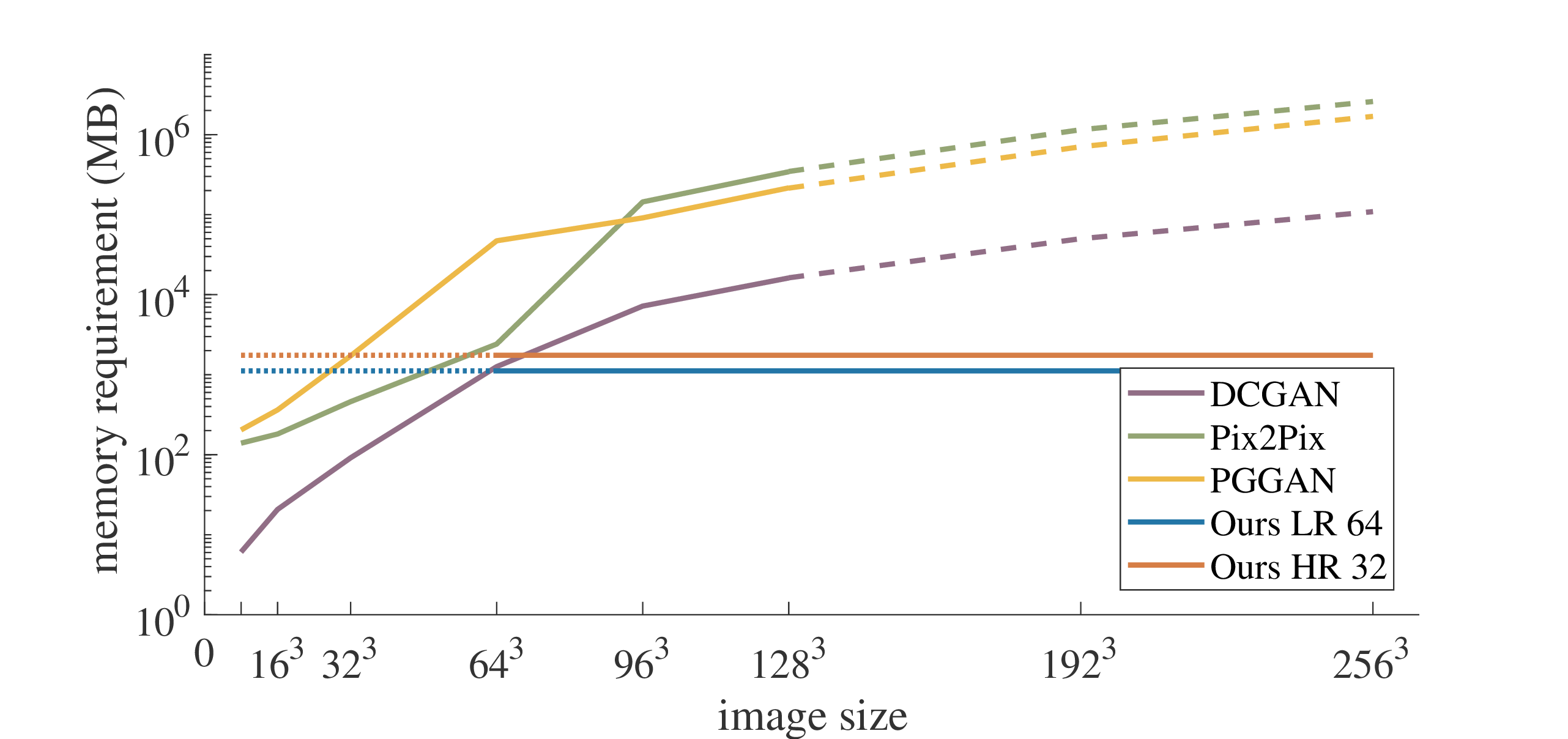}
     \caption{RAM requirements for 3D GANs. Baselines: DCGAN,  Pix2Pix and PGGAN. Dashed lines indicate cubic regression approximation. Our methods: for low resolution images of size $64^3$ (LR 64) and high resolution patches of size $32^3$ (HR 32), have constant memory requirement regardless the image size. Dotted lines indicate sizes under the assumed minimal size  $64^3$. Log-scale is used on the $y$-axis.}
     \label{fig:memuse}
\end{figure}

 GANs are currently rarely applied to 3D images due to computational constrains, therefore in this experiment we investigate the dependence of 3D image side length and memory requirements. Three common GAN architectures are chosen as baselines: DCGAN \cite{3dgan}, Pix2Pix \cite{pix2pix} and progressive growing GAN (PGGAN) \cite{progressive}, and compared to the two architectures of our method: LR 64 for low resolution images of size $64^3$  and HR 32 for high resolution patches of size $32^3$.
 PyTorch \cite{pytorch} is used as implementation framework of choice for all networks.
 Since Pix2Pix and PGGAN are only implemented for 2D images, a straight-forward translation to 3D is obtained (replacing 2D convolutions by 3D convolutions, etc.). The RAM demand computation is realized using the summary approach from the keras framework \cite{keras}. The assumed lower bound of memory usage here includes one forward and backward pass for the generator and discriminator each, as well as the memory required to store the images, gradients and network's parameters for batch size of one.
 The results for different image sizes are shown in Fig.~\ref{fig:memuse}. Naturally, all three baseline approaches have an at least cubic memory requirement growth w.r.t. the image side's length. For those approaches calculations for size over $128^3$ were not even possible on the used Titan XP 12GB GPU, thus the extrapolated cubic regressed curves are shown. These results underline the infeasibility of straight-forward 3D GAN approaches for medical images, as their sizes commonly reach $512^3$. In contrast, our method has a constant character and is thus suitable for arbitrary image sizes with predictable memory usage.      

\subsection{High Resolution Medical Image Generation}
\begin{figure}[t]
\begin{tabular}{cccc}
\includegraphics[height=0.2\textwidth,width=0.24\textwidth]{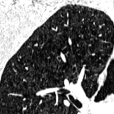} &
\includegraphics[height=0.2\textwidth,width=0.24\textwidth]{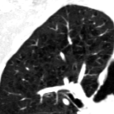} &
\includegraphics[height=0.2\textwidth,width=0.24\textwidth]{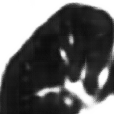} &
\includegraphics[height=0.2\textwidth,width=0.24\textwidth]{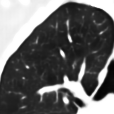} \\
Real B80f image & Real B20f image & Patch B80f $\rightarrow$ B20 & Our B80f $\rightarrow$ B20f
\\
\includegraphics[height=2cm,width=0.2\textwidth]{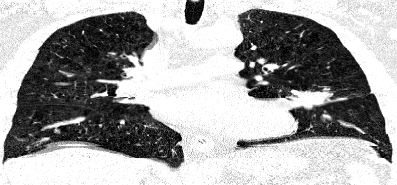} &
\includegraphics[height=2cm,width=0.2\textwidth]{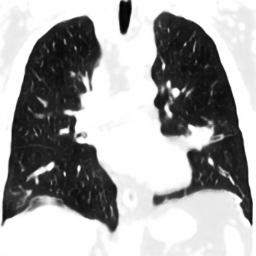} &
\includegraphics[height=2cm,width=0.2\textwidth]{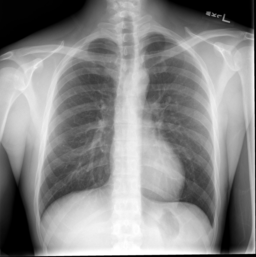} &
\includegraphics[height=2cm,width=0.2\textwidth]{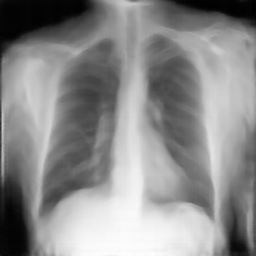} \\
 Real low-dose & Low-dose $\rightarrow$ high-dose& Real X-ray & Generated X-ray
\end{tabular}
\caption{Exemplary images from the used datasets and results of the experiments. First row: thorax CTs (zoomed) -- real B80f image; corresponding B20f; translated B80f to B20f with a standard patch-wise approach; our method. Second row: Real low-dose image; translated low-dose to high-dose; real X-ray; generated X-ray.\protect\footnotemark}
\label{fig:images}
\end{figure}

\paragraph*{\textit{\textbf{Data.}}} Our experiments use the following datasets (see Fig.~\ref{fig:images} for examples):

\textit{Thorax CT:} 3D CT images ($\sim 512^3$) of 56 subjects with varying degree of chronic obstructive pulmonary disease. For each subject the data is reconstructed
simultaneously with a soft (B20f) and a sharp (B80f) kernel. 

\textit{Low-dose thorax CT:} Low-dose 4D lung CT images (120kVp, 40mAs) of 12 patients acquired during free breathing, resulting in 166 3D volumes in total. Image intensities are rescaled with a lung window setting and the images are cropped with a bounding box around the lungs resulting in about $320^3$ voxels.

\textit{Thorax X-ray:} An open access chest X-ray dataset from the Indiana University containing frontal and lateral images usually around the size of $2000^2$. In our experiments about 1500 frontal images of size $2048^2$ are used. 

\noindent\textit{\textbf{Experimental Setup.}}
Our experiments are based on unpaired image domain translation, so the trained GANs are conditioned on the object's edges. A few scenarios are considered:
(1) B80f to B20f CTs: Since images reconstructed with a sharp B80f kernel are very noisy, a reconstruction with a soft B20f kernel is advantageous for automatic quantitative computations (e.g. emphysema index). For this purpose, the GANs  are trained in a 5-fold cross-validation manner on the B20f images from the thorax dataset. In test phase, the edges from the B80f images of the test patients are extracted and translated into the B20f domain using the trained GANs. (2) Low-dose to high-dose: In this experiment, we show that translations between different image domains of different devices and protocols are possible with our method, since it does not require paired data. Here, the GANs are trained on the high-dose B20f images from the thorax dataset, and in test phase, the edges extracted from the low-dose data are used to enable a translation between the two datasets. (3) Large 2D image generation: To show that our approach is not constricted to 3D images, in this experiment huge X-ray images are generated. The GANs are trained on $90\%$ of the thorax X-ray datasets, whereas the rest of the images is used for testing.

\noindent\textit{\textbf{Results.}} For evaluation of experiment (1) the actual B20f reconstructions of the test B80f images serve as ground truth and for experiment (3) the real X-ray images. Here structural similarity index (SSIM), mean absolute error (MAE) and mean squared error (MSE) between the ground truth and the generated images are used as image quality criterions (Tab.~\ref{tab:results}). Our method is compared to other
standard methods: generate a smaller image with an LR GAN ($64^3$ is about our computational limit) and upscale it; develop a straight-forward patch-based approach for the high resolution images (HR GAN without conditioning on previous scales) and apply stitching.  The comparison networks were trained in the exact same manner as for our method. In all experiments isotropic patch sizes of 64 for the LR GAN, and 32 for the HR GANs are used. To avoid padding artifacts on patch borders, only the network's receptive field parts of the generated patches are considered. However, for our method no patch overlaps are needed, whereas for the conventional patch-based method, patch overlaps of 5 pixels are applied. 
The significantly higher SSIM values and smaller pixel-wise distances for our method show its ability to outperform conventional GAN-based approaches. Visually, the generated images have a consistent appearance, barely any artifacts and high resolution, enabling the visibility of tiny structures within the lungs. In contrast, when using conventional patch-wise approaches, patch artifacts and inconsistencies are clearly present (Fig.~\ref{fig:images}). 

We further apply a state-of-the-art denoising method, the non-local means filter, to match the B80f and B20f images. In terms of metrics, this approach is comparable to ours, however its computation time is $\sim100$ times longer and it does not apply to any other domain transfer tasks. Also, the filtered images lack important details and have an inconsistent appearance visually. \footnotemark 

For the low-dose dataset no ground truth is available, so we evaluate experiment (2) qualitatively and compare the feature space distributions of both domains using the Fr\'echet inception distance (FID) between: the original low-dose and the original high-dose images (150); the translated and the original high-dose images (131); the translated and the original low-dose images (157). The visual correspondence of the target domain and the translated images is also underlined by the smaller FID value. The large FID between the original low-dose and the translated images indicates that different features are extracted for the same image of various appearances, and thus emphasizes the need for domain translation.
\setcounter{footnote}{0}
\footnotetext{See supplementary for larger/ further images}
\begin{table}[h!]
\caption{Quantitative results: Measurements between a generated image and its ground truth. Columns 3-5: average SSIM (higher is better), MAE and MSE (lower is better). Experiments (top and bottom): B80f to B20f image translation on a thorax CT dataset, and image generation on thorax X-Ray. Compared to ground truth (row-wise): For both experiments -- our generated images, upscaled low resolution images, conventional patch-wise generation; for thorax CT -- the original B80f image and a non-local means filtered B80f image. Subscript ($^\star$) indicates significantly worse performing methods in terms of all measures compared to ours in a paired t-test ($p<0.0001$).}
\label{tab:results}
\centering\footnotesize
\begin{tabular}{l@{\hspace{0.5em}} l@{\hspace{0.5em}}c@{\hspace{1em}}c@{\hspace{1em}}c}
\footnotesize
               & & \textbf{SSIM} & \textbf{MAE} & \textbf{MSE}\\
\textbf{Dataset}& \textbf{Method} & mean($\pm std$) & mean($\pm std$) & mean($\pm std$)
\\\hline
\multirow{5}{*}{\begin{tabular}{@{}c@{}}Thorax CT \\ $512\times 512 \times 512$ \end{tabular} } & Our gen. B20f &$\mathbf{0.773\pm 0.025}$&$0.033\pm 0.004$&$\mathbf{0.004\pm 0.001}$\\
                            & Small gen. B20f$\mathbf{^\star}$ &$0.633\pm 0.024$&$0.058\pm 0.004$&$0.011\pm 0.002$\\
                            & Patch gen. B20f$\mathbf{^\star}$ &$0.706\pm 0.047$&$0.049\pm 0.007$&$0.008\pm 0.002$\\
                             & Original B80f$\mathbf{^\star}$ &$0.480\pm 0.045$&$0.065\pm 0.008$&$0.012\pm 0.003$\\
                           & Filtered B80f &$\mathbf{0.773\pm 0.048}$&$\mathbf{0.031\pm 0.005}$& $\mathbf{0.004\pm 0.001}$\\
\hline

\multirow{3}{*}{\begin{tabular}{@{}c@{}}Thorax X-Ray \\ $2048 \times 2048$ \end{tabular}} & Our gen. &$\mathbf{0.711\pm 0.067}$&$\mathbf{0.104\pm 0.028}$&$\mathbf{0.022\pm 0.014}$\\
                            & Small gen.$\mathbf{^\star}$ &$0.673\pm 0.064$&$0.108\pm 0.027$&$0.024\pm 0.014$\\
                            & Patch gen.$\mathbf{^\star}$ &$0.520\pm 0.072$&$0.200\pm 0.024$&$0.069\pm 0.019$\\
\hline

\end{tabular}
\end{table}
\section{Discussion and Conclusion}\label{sec:discussion}
In this work, we propose a multi-scale GAN-based approach for the generation of arbitrarily large 3D medical images of high resolution and realistic homogeneous appearance. We show that in contrast to existing GANs, that produce the whole image at once, the presented patch-based method, only requires constant GPU memory w.r.t. the image size. Also compared to trivial patch-wise methods, our sophisticated multi-resolution scheme provides higher quality images in terms of image consistency and resolution.  Here we show a use-case scenario for medical image domain translation based on conditional GANs, however our method is also suitable for image generation from scratch and it enables various applications and possibilities for medical images.

%
%
%
%
\bibliographystyle{splncs04}
\bibliography{3DGAN}
\appendix
\section{Supplementary}
\subsection{Network Architectures}
\begin{figure}[h!]
    \centering
    \includegraphics[width=0.9\textwidth]{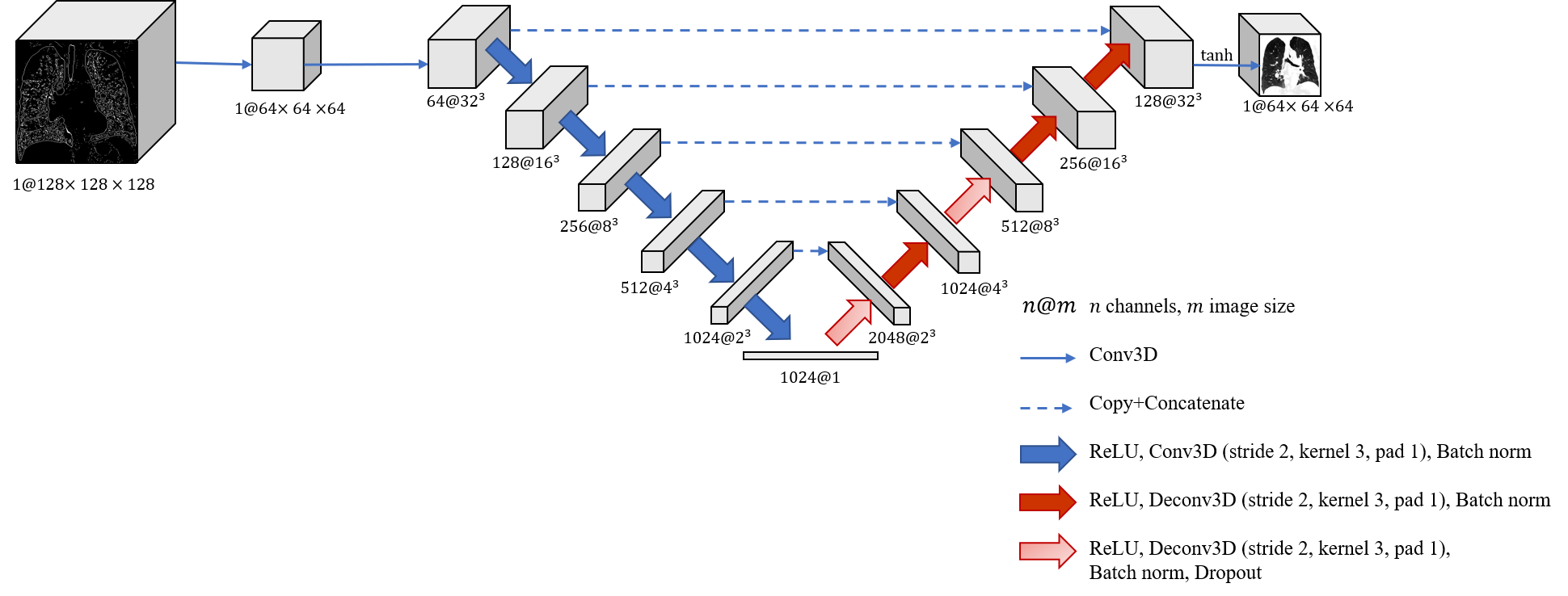}
    \caption{LR generator architecture}
    \label{fig:LRArch}
\end{figure}
\begin{figure}[h!]
    \centering
    \includegraphics[width=0.9\textwidth]{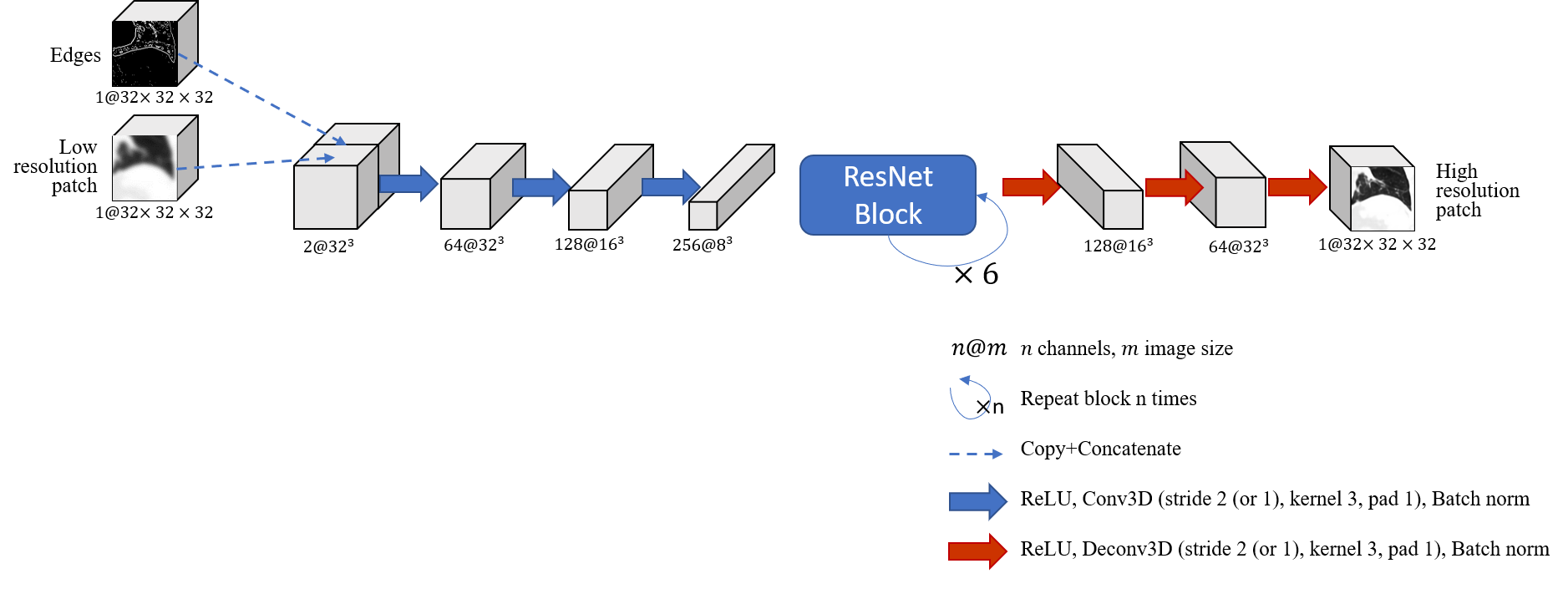}
    \caption{HR generator architecture}
    \label{fig:HRArch}
\end{figure}
\begin{figure}[h!]
    \centering
    \includegraphics[width=0.9\textwidth]{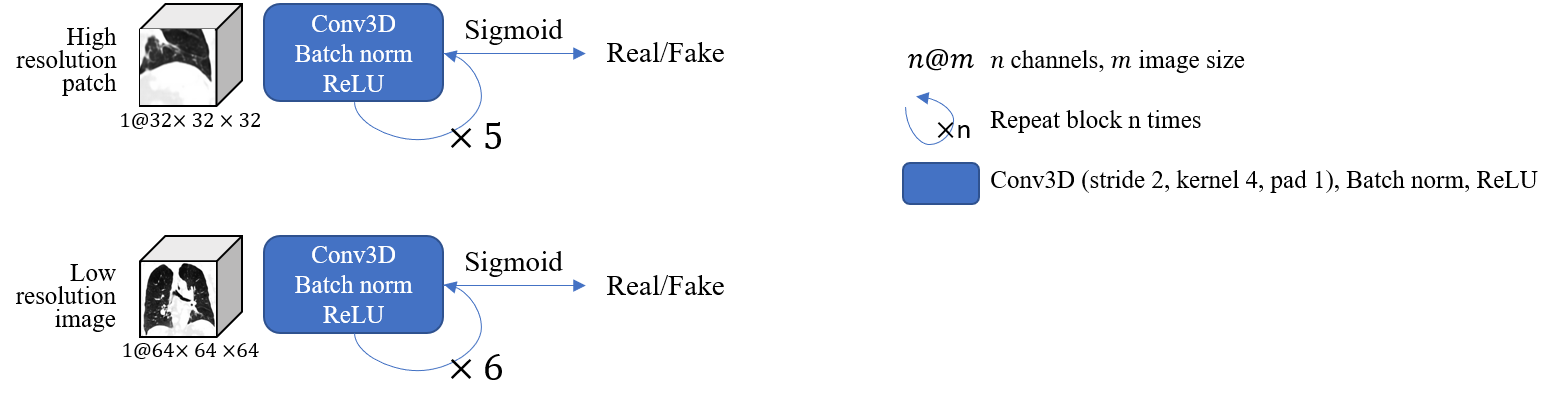}
    \caption{LR and HR networks discriminator architectures}
    \label{fig:DArch}
\end{figure}

\subsection{3D thorax CT example image slices (axial, coronal,sagittal)}

\begin{figure}[h!]
\centering
\begin{tabular}{cc}
\multicolumn{2}{c}{Example 1: Sharp (B80f) to soft (B20f) kernel scenario, image size $512^3$}\\
Real B80f  & Real B20f \\
\includegraphics[height=0.45\textwidth,width=0.45\textwidth]{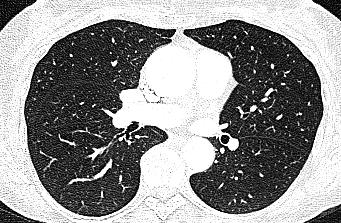} &
\includegraphics[height=0.45\textwidth,width=0.45\textwidth]{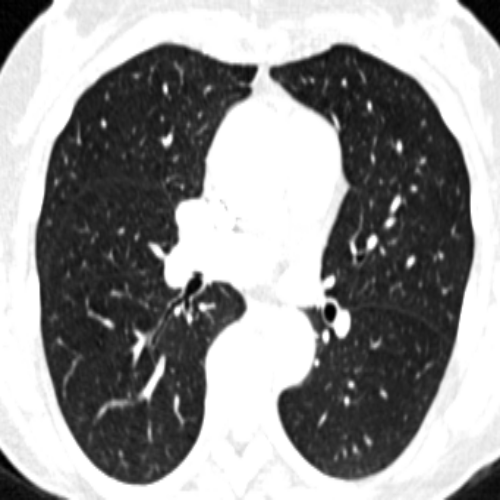}
\\
\includegraphics[height=0.45\textwidth,width=0.45\textwidth]{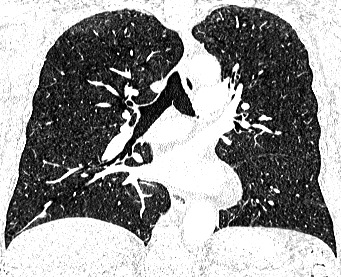} &
\includegraphics[height=0.45\textwidth,width=0.45\textwidth]{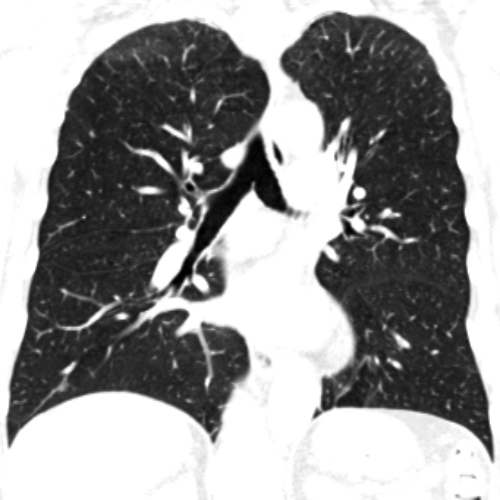}
\\
\includegraphics[height=0.45\textwidth,width=0.45\textwidth]{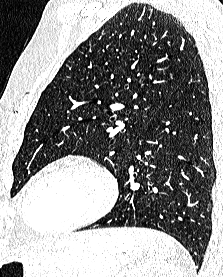} &
\includegraphics[height=0.45\textwidth,width=0.45\textwidth]{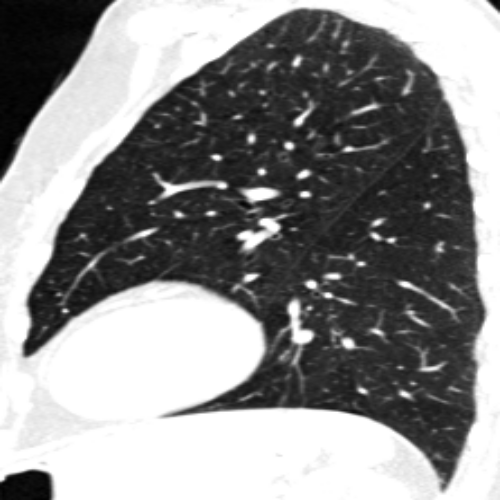}
\\
\end{tabular}
\end{figure}

\newpage

\begin{figure}[h!]
\centering
\begin{tabular}{cc}
Fake B80f$\rightarrow$B20f, ours & Fake B80f$\rightarrow$B20f, patchwise \\

\includegraphics[height=0.45\textwidth,width=0.45\textwidth]{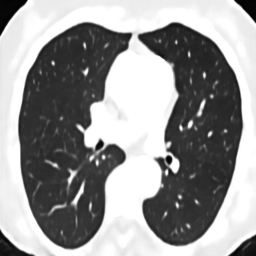} &
\includegraphics[width=0.45\textwidth]{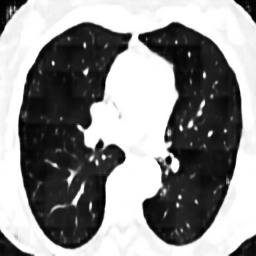}
\\
\includegraphics[height=0.45\textwidth,width=0.45\textwidth]{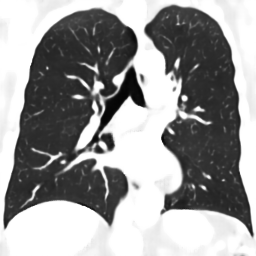} &
\includegraphics[width=0.45\textwidth]{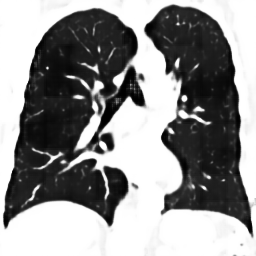}
\\
\includegraphics[height=0.45\textwidth,width=0.45\textwidth]{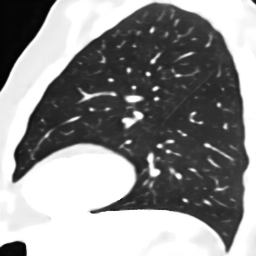} &
\includegraphics[width=0.45\textwidth]{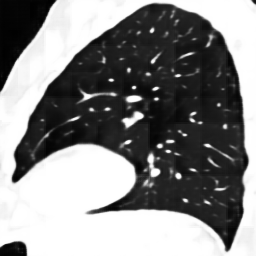}
\\
\end{tabular}
\end{figure}

\newpage

\begin{figure}[h!]
\centering
\begin{tabular}{cc}
Non-local means filtered B80f & \\

\includegraphics[height=0.45\textwidth,width=0.45\textwidth]{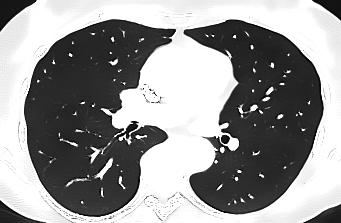} &
\\
\includegraphics[height=0.45\textwidth,width=0.45\textwidth]{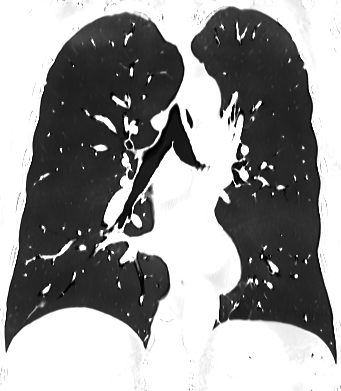} &
\\
\includegraphics[height=0.45\textwidth,width=0.45\textwidth]{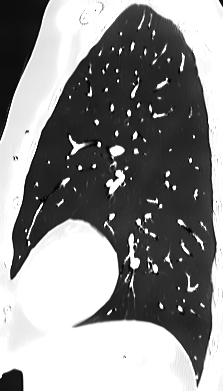} &
\\
\end{tabular}
\end{figure}

\newpage

\begin{figure}[h!]
\centering
\begin{tabular}{cc}
\multicolumn{2}{c}{Example 2: Sharp (B80f) to soft (B20f) kernel scenario, image size $512^3$}\\
Real B80f  & Real B20f \\
\includegraphics[height=0.45\textwidth,width=0.45\textwidth]{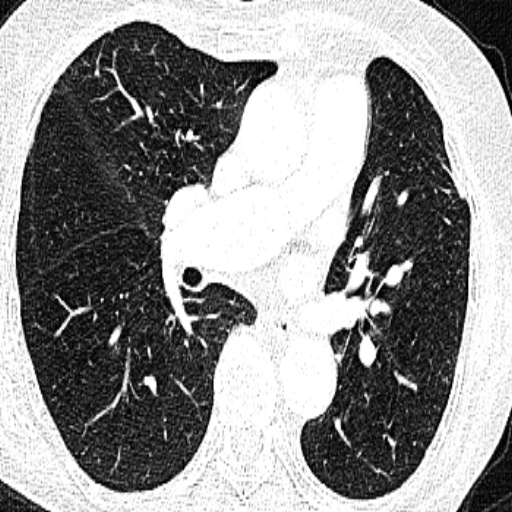} &
\includegraphics[width=0.45\textwidth]{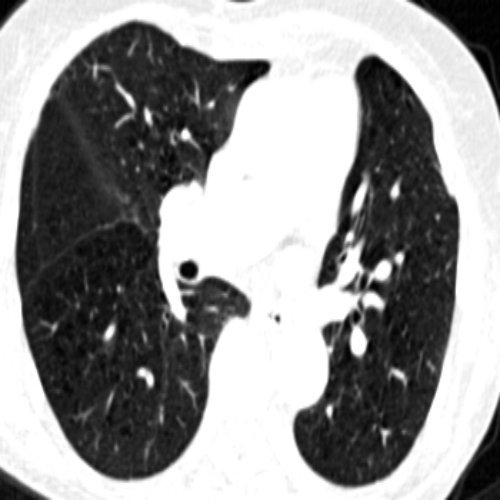}
\\
\includegraphics[height=0.45\textwidth,width=0.45\textwidth]{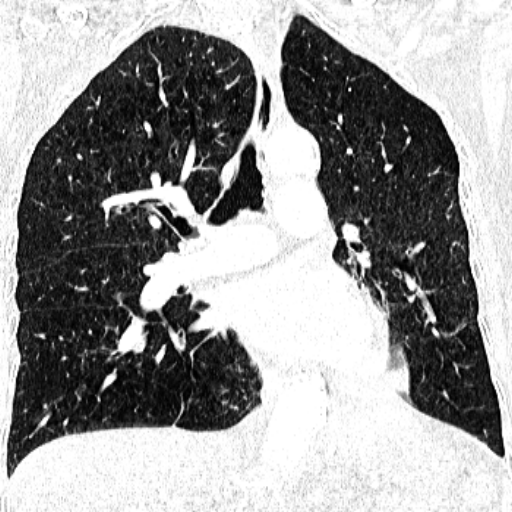} &
\includegraphics[width=0.45\textwidth]{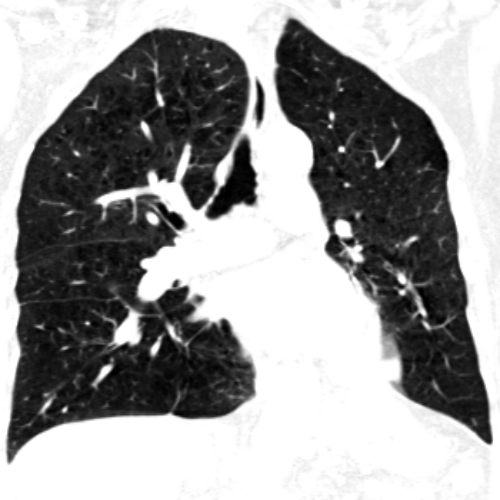}
\\
\includegraphics[height=0.45\textwidth,width=0.45\textwidth]{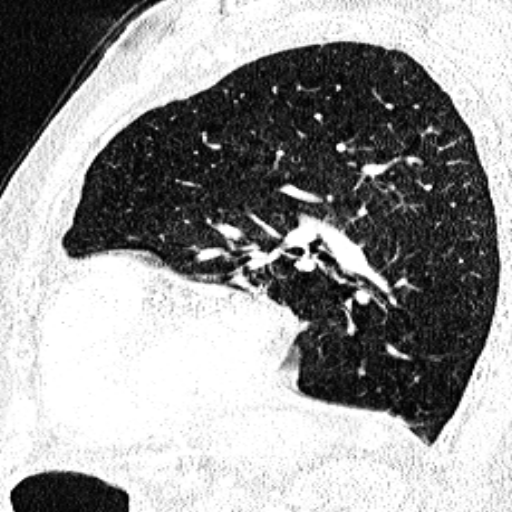} &
\includegraphics[width=0.45\textwidth]{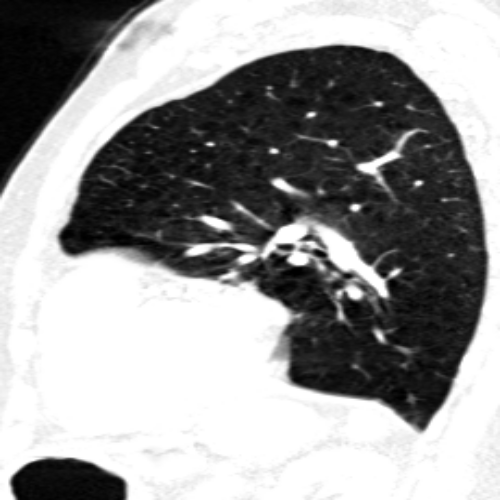}
\\
\end{tabular}
\end{figure}

\newpage

\begin{figure}[h!]
\centering
\begin{tabular}{cc}
Fake B80f$\rightarrow$B20f, ours & Fake B80f$\rightarrow$B20f, patchwise \\

\includegraphics[height=0.45\textwidth,width=0.45\textwidth]{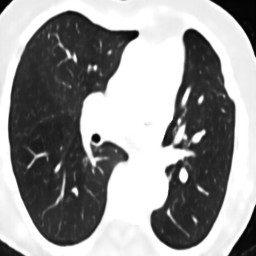} &
\includegraphics[width=0.45\textwidth]{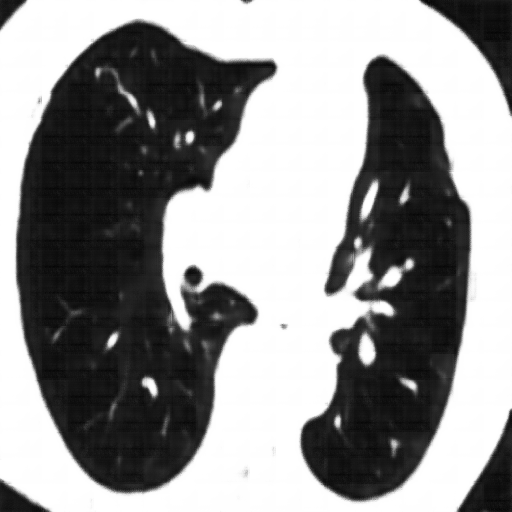}
\\
\includegraphics[height=0.45\textwidth,width=0.45\textwidth]{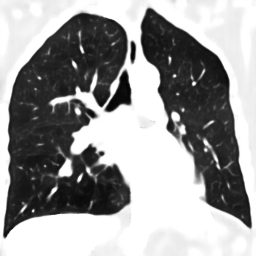} &
\includegraphics[width=0.45\textwidth]{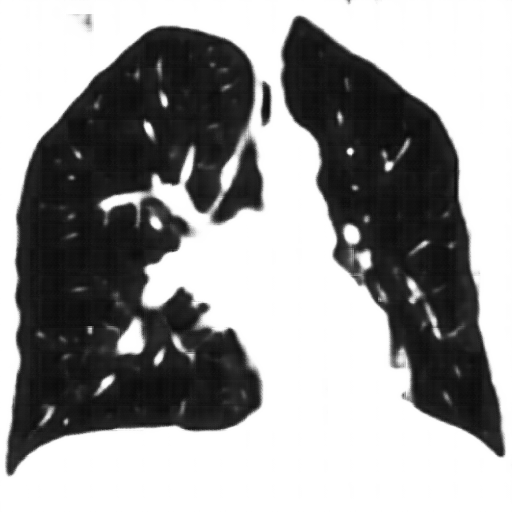}
\\
\includegraphics[height=0.45\textwidth,width=0.45\textwidth]{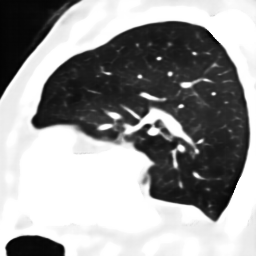} &
\includegraphics[width=0.45\textwidth]{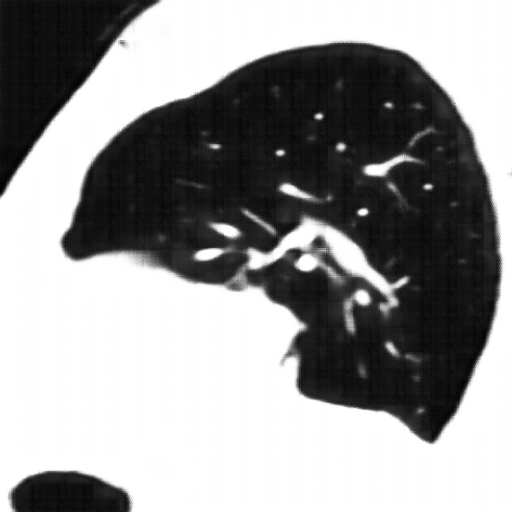}
\\
\end{tabular}
\end{figure}

\newpage

\begin{figure}[h!]
\centering
\begin{tabular}{cc}
Non-local means filtered B80f & \\

\includegraphics[height=0.45\textwidth,width=0.45\textwidth]{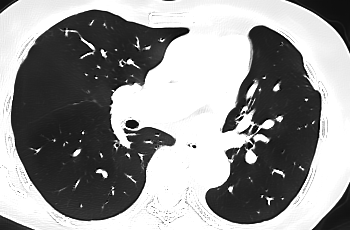} &
\\
\includegraphics[height=0.45\textwidth,width=0.45\textwidth]{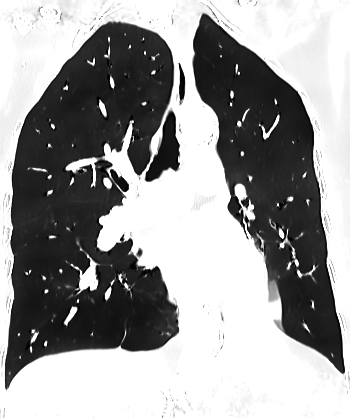} &
\\
\includegraphics[height=0.45\textwidth,width=0.45\textwidth]{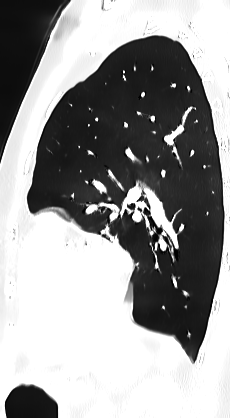} &
\\
\end{tabular}
\end{figure}

\newpage

\begin{figure}[h!]
\centering
\begin{tabular}{cc}
\multicolumn{2}{c}{Example 1: Low-dose to high-dose scenario, image size $256^3$}\\
Real low-dose  & Fake high-dose, our method \\
\includegraphics[height=0.45\textwidth,width=0.45\textwidth]{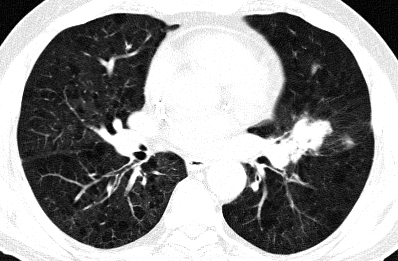} &
\includegraphics[width=0.45\textwidth]{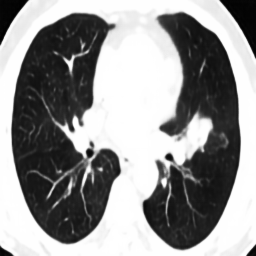}
\\
\includegraphics[height=0.45\textwidth,width=0.45\textwidth]{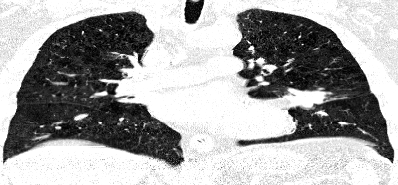} &
\includegraphics[width=0.45\textwidth]{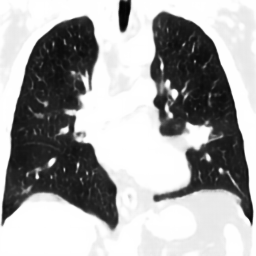}
\\
\includegraphics[height=0.45\textwidth,width=0.45\textwidth]{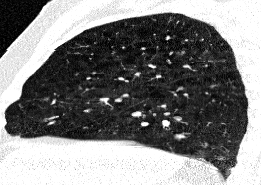} &
\includegraphics[width=0.45\textwidth]{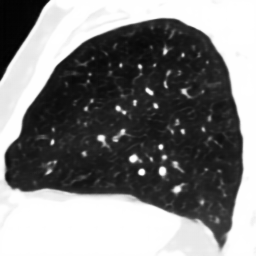}
\\
\end{tabular}
\end{figure}

\newpage

\begin{figure}[h!]
\centering
\begin{tabular}{cc}
\multicolumn{2}{c}{Example 2: Low-dose to high-dose scenario, image size $256^3$}\\
Real low-dose  & Fake high-dose, our method \\
\includegraphics[height=0.45\textwidth,width=0.45\textwidth]{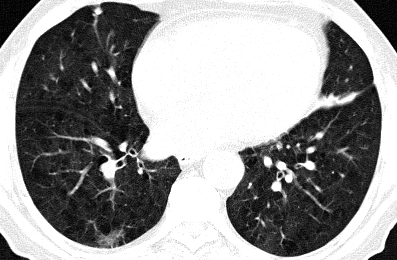} &
\includegraphics[width=0.45\textwidth]{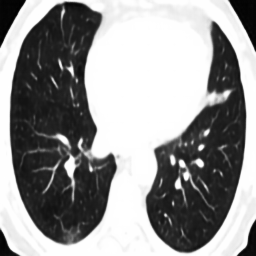}
\\
\includegraphics[height=0.45\textwidth,width=0.45\textwidth]{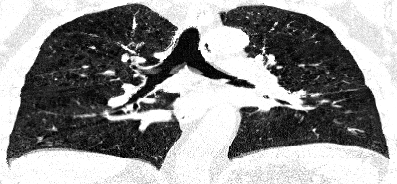} &
\includegraphics[width=0.45\textwidth]{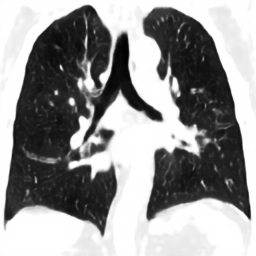}
\\
\includegraphics[height=0.45\textwidth,width=0.45\textwidth]{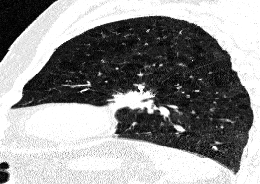} &
\includegraphics[width=0.45\textwidth]{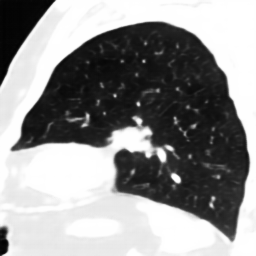}
\\
\end{tabular}
\end{figure}

\newpage

\subsection{3D brain MRI example image slices (axial): T1 to T2 translation}

\begin{figure}[h!]
\centering
\begin{tabular}{cccc}
\multicolumn{3}{c}{Examples: T1 to T2 scenario, image size $240\times 240\times 155$}\\
Real T1 (GT) & Sketch (Input) & Real T2 & Fake T2 \\
\includegraphics[width=0.24\textwidth]{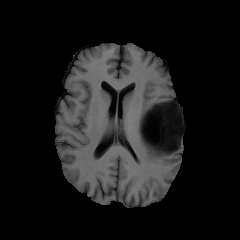} &
\includegraphics[width=0.24\textwidth]{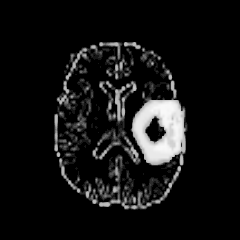} &
\includegraphics[width=0.24\textwidth]{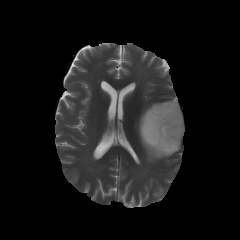}
&
\includegraphics[width=0.24\textwidth]{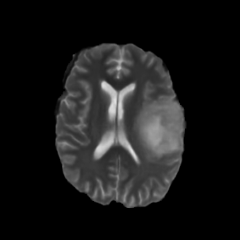}
\\
\includegraphics[width=0.24\textwidth]{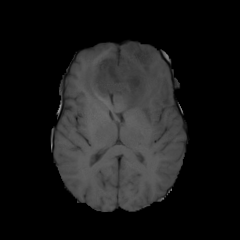} &
\includegraphics[width=0.24\textwidth]{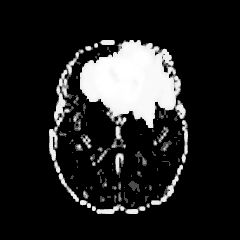} &
\includegraphics[width=0.24\textwidth]{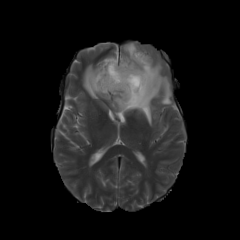}
&
\includegraphics[width=0.24\textwidth]{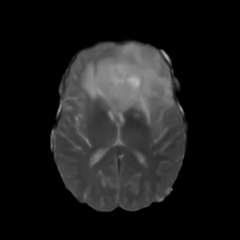}
\\
\includegraphics[width=0.24\textwidth]{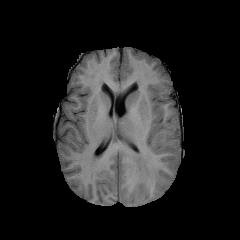} &
\includegraphics[width=0.24\textwidth]{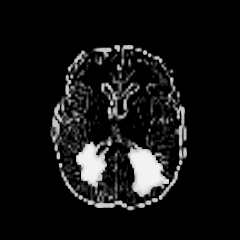} &
\includegraphics[width=0.24\textwidth]{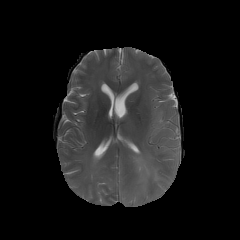}
&
\includegraphics[width=0.24\textwidth]{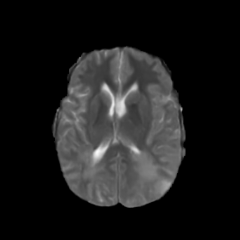}
\\
\end{tabular}
\end{figure}

\newpage
\subsection{2D thorax X-ray example images}
\begin{figure}[h!]
\centering
\begin{tabular}{ccc}
\multicolumn{3}{c}{Examples: generated X-ray images, image size $2048\times 2048$}\\
\includegraphics[width=0.33\textwidth]{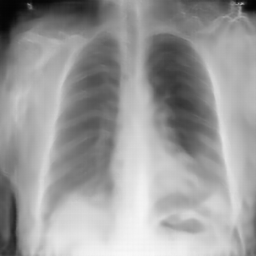} &
\includegraphics[width=0.33\textwidth]{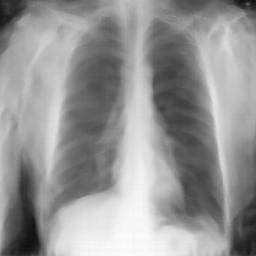}
&
\includegraphics[width=0.33\textwidth]{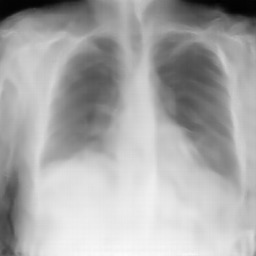}
\\
\includegraphics[width=0.33\textwidth]{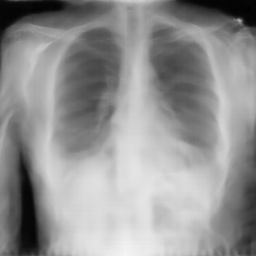} &
\includegraphics[width=0.33\textwidth]{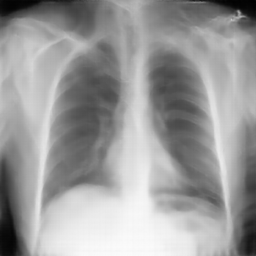}
&
\includegraphics[width=0.33\textwidth]{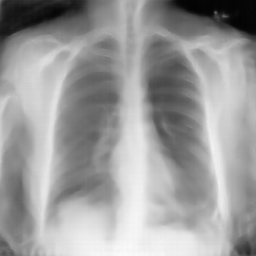}
\\
\includegraphics[width=0.33\textwidth]{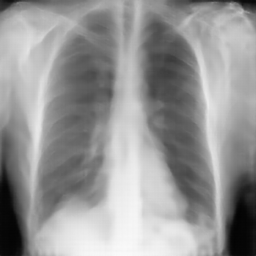} &
\includegraphics[width=0.33\textwidth]{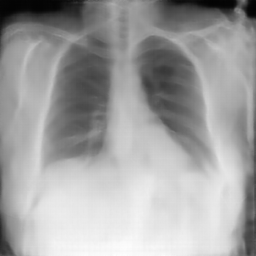}
&
\includegraphics[width=0.33\textwidth]{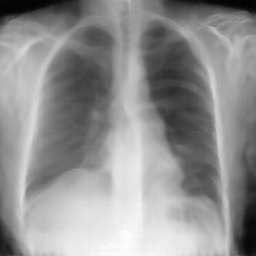}
\\
\end{tabular}
\end{figure}

\end{document}